\documentclass[3p,times,procedia]{elsarticle}
\usepackage{nupha_ecrc}

\volume{00}

\firstpage{1}

\journalname{Nuclear Physics A}

\runauth{}

\jid{nupha}

\jnltitlelogo{Nuclear Physics A}

\usepackage{graphicx}
\usepackage{latexsym}
\usepackage{amsfonts}
\usepackage{amssymb}
\usepackage{amsmath}
\usepackage{bm}
\usepackage{mathrsfs}
\usepackage{slashed}

\newcommand{\bk}{{\bm{k}}}

\newcommand{\para}{ \parallel}

\usepackage[normalem]{ulem}  

\usepackage{color} 

\renewcommand\sout{\bgroup \color{red} \ULdepth=-.5ex \ULset}

\usepackage[figuresright]{rotating}

\begin{document}

\begin{frontmatter}





\dochead{XXVIIth International Conference on Ultrarelativistic Nucleus-Nucleus Collisions\\ (Quark Matter 2018)}

\title{Transport phenomena with chiral fermions 
\\
in strong magnetic fields}


\author[Fudan]{Koichi Hattori}

%
%
%
%
%
%

\address[Fudan]{Physics Department and Center for Particle Physics and Field Theory,
Fudan University, Shanghai 200433, China}
%
%
%
%
%
%
%

\begin{abstract}
I discuss the electrical conductivity, bulk viscosity, and heavy-quark diffusion coefficient 
in strong magnetic fields putting emphasis on consequences of the chirality conservation 
on the perturbative relaxation dynamics. 
\end{abstract}


\end{frontmatter}

\section{Introduction}


The transport phenomena induced by chiral anomaly and 
the magnetic field in the relativistic heavy-ion collisions 
have attracted a number of interestes in the last decade (see other contributions to Quark Matter 2018 and, e.g., Ref.~\cite{Hattori:2016emy} for a review). 
Many aspects of these phenomena can be understood from 
the special properties of the chiral fermions in the lowest Landau levels (LLLs) 
which have the (1+1)-dimensional linear dispersion relations. 
In this contribution, I discuss consequences of such properties reflected 
in more conventional transport phenomena, i.e., 
electrical conductivity~\cite{Hattori:2016lqx,Hattori:2016cnt}, bulk viscosity~\cite{Hattori:2017qih}, 
and heavy-quark diffusion coefficient~\cite{Fukushima:2015wck}, 
when the magnetic field strength is much larger than the temperature scale. 


\section{Dissipative transport phenomena in strong magnetic fields}

\subsection{Longitudinal electrical conductivity and bulk viscosity}


In the presence of a preferred spatial orientation provided by a magnetic field, 
there are three and seven components of electrical conductivities and 
viscosities, respectively~\cite{Huang:2011dc}. 
However, 
the fermion in the LLL can transport the (electric) charge and energy-momentum only 
along the magnetic field $ [ {\bm B} = (0,0,B)] $. 
Consequently, the electric current $ j^\mu $ and the energy-momentum tensor $ T^{\mu\nu} $ 
have only the temporal and longitudinal components. 
(More precisely, this is true for $ T^{\mu\nu} (q)$ 
when the transverse momenta $ q_{x,y} $ are sent to zero~\cite{Hattori:2017qih}.)
Therefore, the quark carriers 
contribute only to the longitudinal current $ j_z $ and longitudinal pressure $  P_\para   $, 
corresponding to the longitudinal conductivity $  \sigma_{zz}$ and bulk viscosity $ \zeta_\para $, respectively.

In the linear response regime, the longitudinal transport coefficients 
$  \sigma_{zz}$ and $ \zeta_\para $ at $ T \not=0$ are given by 
\begin{subequations}
\begin{eqnarray}
&&
\sigma_{zz} = \frac{ j_z}{E_z} = N_c \sum_{f} e_f \frac{| e_f B |}{ 2\pi } \cdot 
2 \int_{-\infty}^{\infty}  \frac{dp_z}{2\pi} \frac{p_z}{\epsilon_{p} } \left[ \frac{\delta f(p_z)}{E_z} \right]
\label{eq:conductivity}
\, ,
\\
&&
\zeta_\para = - \frac{1}{3}   \frac{ \tilde P_{\para} }{\partial _z u_z} 
= - \frac{N_c}{3} \sum_{f}   \frac{| e_f B |}{ 2\pi } \cdot 
2 \int_{-\infty}^{\infty}  \frac{dp_z}{2\pi}  \frac{p_z ^2 -  \Theta_\beta  \epsilon_{p} ^2}{ \epsilon_{p} }
\left[  \frac{ \delta f(p_z) }{\partial _z u_z} \right]
\label{eq:bulk-viscosity}
\, ,
\end{eqnarray}
\end{subequations}
where $ N_c $ and $ e_f $ are the number of colors and 
the electric charge carried by the $ f $-flavored quark, respectively. 
The two-dimensional transverse phase space is degenerated 
with the density of states $ | e_f B | / (2\pi)  $, 
corresponding to the translational invariance for 
the center coordinates of the cyclotron motions. 
To obtain the off-equilibrium component of the pressure $   \tilde P_{\para}$, 
one needs to subtract the equilibrium component with 
$  \Theta_\beta = (\partial P_\para/\partial \epsilon)_B$, 
because the pressure decreases even in an adiabatic expansion/compression~\cite{Arnold:2006fz}. 
In the massless limit, the adiabatic expansion rate is given by 
$  \Theta_\beta 
= 1$ that is the inverse of the number of spatial dimensions. 
Clearly, the bulk viscosity vanishes in the massless limit 
where the scale invariance preserves the equilibrium state at any step of the expansion/compression. 
Therefore, the bulk viscosity is sensitive to the quark mass that breaks the scale invariance. 
In the massive case, we have $  \Theta_\beta = 1 - 3m^2/(\pi^2 T^2)$~\cite{Hattori:2017qih}.

The off-equilibrium components $ \delta f(p_z) = f(p_z)-f_{\rm eq}(p_z)$ 
of the particle distribution functions $ f(p_z) $ will be obtained as solutions of the Boltzmann equations. 
The relaxation dynamics in the LLL is described by a $(1+1)$-dimensional Boltzmann equation, 
since the transverse momentum only servers as a label of 
the degenerated states. 
In the presence of an electric field $ E_z $ and an expansion/compression flow $u_z (z) $, 
the explicit forms of the Boltzmann equations are, respectively, given by  
\begin{subequations}
\begin{eqnarray}
\label{eq:Boltzmann1}
&& 
e_f E_z \frac{ \partial f (p_z)}{\partial_{p_z}}  = C[f]
\, ,
 \\
&&
\label{eq:Boltzmann2}
(\partial_t +v_z \partial_z ) f(p_z; t,z) = C[f]
\, ,
\end{eqnarray}
\end{subequations}
where $v_z \equiv p_z/\epsilon_p$ is the velocity in the direction of $B$. 
We have taken the steady and homogeneous limits in Eq.~(\ref{eq:Boltzmann1}). 
The equilibrium distribution function 
is given by $f_{\text{eq}}(p_z) = [ \exp (\beta \epsilon_p) +1 ]^{-1}$ without the flow velocity 
and by $f_{\text{eq}}(p_z,t,z) = [ \exp\{\beta(t)\gamma_u (\epsilon_p-p_z u_z) \}+1]^{-1}$ 
in the presence of the flow $u_z$ with $\gamma_u = (1-u_z^2)^{-1/2}$ being the gamma factor. 
In the latter case, the temperature depends on time, 
since the energy density decreases during the adiabatic expansion/compression 
of the system~\cite{Arnold:2006fz}.

The electrical and momentum currents reach steady states 
when the (external) driving force and the collisional effects are balanced. 
To compute the collision term $C[f]$, 
I use perturbation theory with respect to the QCD coupling constant 
on the basis of appropriate resummations in finite temperature and a strong magnetic field. 
As one can imagine from the occurrence of cyclotron radiation, 
the leading-order contributions come from the 1-to-2 (2-to-1) processes. 
On the other hand, the ordinary perturbative expansion, without an external magnetic field, 
starts from the 2-to-2 processes, since the kinematics of the 1-to-2 processes, 
i.e, conversions among two fermions and a massless gauge boson, cannot be satisfied. 
This difference in the kinematics can be understood from the fact that 
the gluons have the normal (3+1)-dimensional dispersion relations without 
being subject to the Landau-level discretization, 
while the quarks have the (1+1)-dimensional ones. 
Because of this mismatch in the dimensions, 
the gluon transverse momentum $ |\bk_\perp|^2 = k_0^2-k_z^2 $ 
can be regarded as an ``effective mass'' in the (1+1)-dimensional kinematics. 
One may be then convinced that the kinematics of the 1-to-2 processes 
with a ``massive gauge boson'' is satisfied~\cite{Hattori:2016lqx, Hattori:2016cnt}.

\begin{figure*}
		\vspace{-0.3cm}
		\begin{center}
			\includegraphics[width=0.52\hsize]{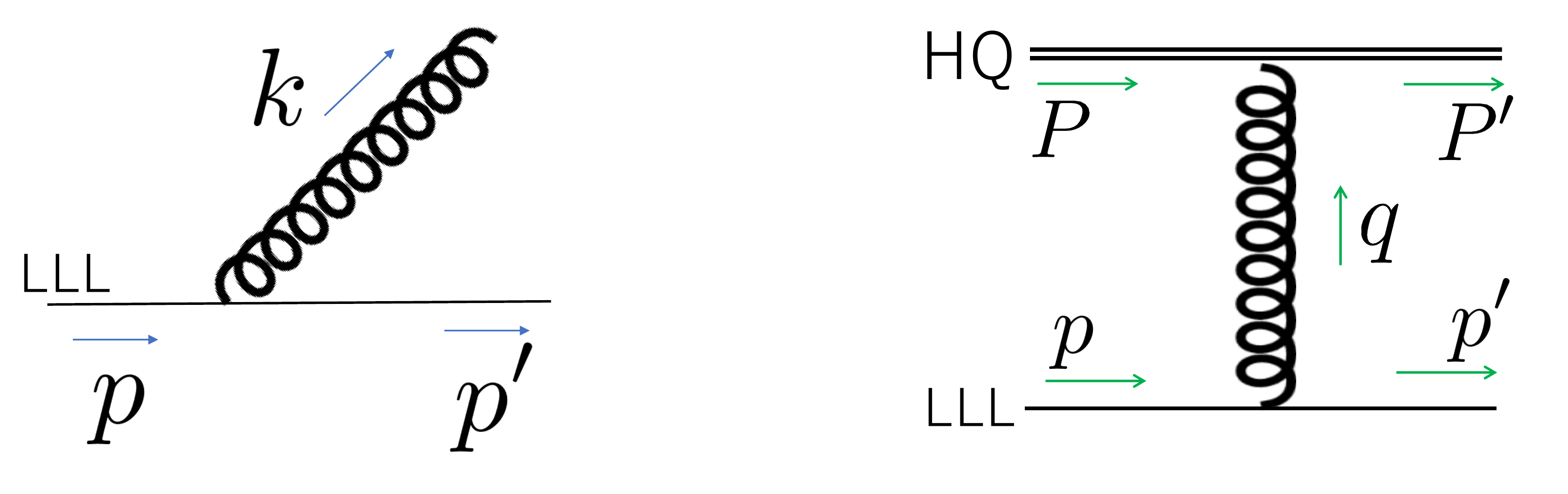}
		\end{center}
		\vspace{-0.8cm}
\caption{An example of the 1-to-2 processes (left) and the one-gluon exchange 
between a heavy quark (HQ) and a thermal light quark (left) in strong magnetic fields. 
The chirality conservation at the vertices plays crucial roles in the relaxation dynamics. }
		\label{fig:diagrams}
\end{figure*}

The massless LLL quarks have the linear dispersion relations $ p^0 = \pm p_z $ 
with the upper (lower) sign for the right- (left-) handed chirality. 
The gluon transverse momentum, or the effective mass, 
can be written by the fermion momenta $p, p^\prime $ involved in the 1-to-2 processes as 
$ |\bk_\perp|^2 = \{ \pm(p_z - p_z^\prime) \}^2 - (p_z - p_z^\prime)^2 
= 0$ (cf., Fig.~\ref{fig:diagrams}).\footnote{
The same conclusion can be drawn for all of the 1-to-2 and 2-to-1 processes. 
} 
This means that the kinematics is satisfied only in the collinear limit $ (k \para p \para p^\prime )  $, 
in which the coupling between a physical transverse gluon and the fermions vanishes. 
Therefore, the collision term is proportional to the square of the current quark mass $  m_f^2$.  
This is a consequence of the chirality conservation at the perturbative interaction vertex, 
which forces the velocities $ p_z/p^0 $ and $ p_z^\prime/p^{\prime \, 0} $ to take the same sign. 
A finite $ m_f $ allows for the chirality mixing.


The Boltzmann equation with the collision term can be solved 
by linearlizing it with respect to the small perturbation. 
Inserting the solution $ \delta f(p_z) $ into Eqs.~(\ref{eq:conductivity}) and (\ref{eq:bulk-viscosity}), 
one obtains the electrical conductivity and the bulk viscosity\footnote{The gluon contribution 
is suppressed by a factor of $ 1/|e_f B| $, because of the large phase space volume of the quark scatterers. 
}~\cite{Hattori:2016lqx,Hattori:2016cnt,Hattori:2017qih} 
\begin{subequations}
\begin{eqnarray}
&&
\sigma_{zz} = N_c \sum_{f} e_f^2 \frac{ \vert e_f B \vert}{2\pi} \frac{4T}{g^2 C_F m_f^2 \ln(T/M)}
\label{eq:sol-conductivity}
\, ,
\\
&&
\zeta_\para = N_c \sum_f  \frac{ \vert e_f B \vert}{2\pi} \frac{m_f^2}{g^2 C_F  \ln(T/M)}
\left[ \frac{4}{\pi^2} - \frac{56}{3} \times (0.0304 \cdots) \right]
\label{eq:sol-bulk-viscosity}
\, ,
\end{eqnarray}
\end{subequations}
where $ C_F = (N_c^2-1)/(2N_c) $. 
The logarithmic factor originates from the collision integral 
with the ultraviolet and infrared cutoff scales at $  T$ and 
$M^2 = {\rm min} [ m_f^2, g^2/(2\pi) \sum_f \vert e_f B \vert/(2\pi) ] $\footnote{
The latter is the gluon screening mass from the quark loop 
which is larger than the gluon-loop contribution when $ eB \gg T^2  $.}, respectively. 
These transport coefficients are enhanced by a factor of $ 1/m_f^2 $ 
as a consequence of the chirality conservation when $ m_f $ is small. 
The bulk viscosity is suppressed by $  m_f^4$ when $ eB =0 $~\cite{Arnold:2006fz}, 
of which the power is reduced to $  m_f^2$ as a consequence of 
the competition between the conformal symmetry and the chirality conservation 
which govern the behaviors in the massless limit. 
It is remarkable that the dependences on the current quark mass are important 
even in the high temperature limit $ T \gg m_f $, as long as $ T^2 \gg e B $. 

\subsection{Heavy-quark diffusion dynamics}

I also discuss the heavy-quark diffusion dynamics in the quark-gluon plasma 
under a strong magnetic field. 
The heavy quarks are dominantly produced 
in the initial hard scatterings among the partons in the colliding nuclei, 
and thus serve as a good probe of the dynamics in the initial and QGP stages. 
Computing the heavy-quark diffusion coefficient in the presence of a strong magnetic field, 
I show that the momentum diffusion along the magnetic field is highly suppressed, 
and the anisotropic diffusion rate gives rise to anisotropic spectra of 
the open heavy-flavor mesons~\cite{Fukushima:2015wck} (see also Ref.~\cite{Hattori:2016emy} for a brief summary)

When the heavy quark is subject to the random kicks by the thermal scatterers, 
its dynamics is described by the Langevin equations~\cite{Moore:2004tg}: 
\begin{eqnarray}
\label{eq:Langevin}
\frac {d p_z}{ dt}
 = -\eta_\parallel p_z + \xi_z \,,
 \qquad
 {d\bm p_\perp\over dt} =
   -\eta_\perp \, \bm p_\perp + \bm\xi_\perp 
   \, ,
\end{eqnarray}
where we have a set of two equations 
for the directions parallel and perpendicular to the magnetic field. 
The random forces are assumed to be white noises that satisfy 
$  \langle \xi_{z}(t) \xi_z(t')\rangle=\kappa_\parallel \delta(t-t')$ 
and $ \langle \xi_{\perp}^i(t)\xi_\perp^j(t')\rangle=\kappa_\perp
  \delta^{ij}\delta(t-t')  $ for $  i,j=x,y $. 
The momentum diffusion coefficients are defined by
\begin{eqnarray}
&&
\kappa_{\parallel} = \int \!\! {d^3\bm q}\,
  {d\Gamma(\bm q)\over d^3\bm q}\, q_{z}^2 
\,, 
\quad 
 \kappa_{\perp} = {1\over 2}\int \!\! {d^3\bm q}\,
  {d\Gamma(\bm q)\over d^3\bm q}\, \bm q_{\perp}^2
  \, ,
\label{kappa2}
\end{eqnarray}
where $ q$ is the momentum transfer from the thermal scatterers to the heavy quark,
and the static limit ($ q^0 \to 0$) is assumed. 
The leading-order contributions to the momentum transfer rate $d\Gamma(\bm q)/ d^3\bm q$ 
come from the one-gluon exchanges with thermal light-quark and gluon scatterers. 

As in the previous section, the linear dispersion relation of the quark scatterers in the LLL 
gives rise to an important kinematics in the one-gluon exchange~\cite{Fukushima:2015wck}. 
Note that we have $q^0 = \pm (p_z^\prime - p_z) $ with $ p  ,  p^\prime$ being 
the momenta of the quark scatterers (cf., Fig.~\ref{fig:diagrams}). 
Again, as a consequence of the chirality conservation, the signs appear only as the overall ones. 
Therefore, one finds that $ q_z = \pm q^0 \to 0 $ in the static limit, 
meaning that the longitudinal momentum transfer is prohibited in the massless limit. 
The consequence of the chirality conservation suggests a strong anisotropy in the diffusion dynamics.

After straightforward computations, one obtains 
the finite contributions from the one-gluon exchange diagrams 
summarized in Table~\ref{table:diffusion} (up to the overall numerical constants). 
The quark contribution to the transverse component is proportional $ eB $, 
while this factor is replaced by $  T^2$ in the isotropic gluon contribution. 
This difference simply originates from the phase space volumes 
of the quark and gluon scatterers. 
Based on these results, one can estimate the anisotropy as 
\begin{eqnarray}
\frac{ \kappa_\para }{ \kappa_\perp } = 
\frac{ \kappa _{\rm gluon} }{ \kappa_{\rm quark} + \kappa _{\rm gluon}  }
\sim \frac{T^2}{eB} 
\, ,
\end{eqnarray}
where $ \kappa_{\rm quark} = \kappa_\perp^{\rm quark} $ and 
$ \kappa _{\rm gluon} =  \kappa _\para^{\rm gluon} =  \kappa _\perp^{\rm gluon} $. 
The diffusion coefficients $\kappa_{\parallel,\perp}$, are
related to the drag coefficients $\eta_{\parallel, _\perp}$ 
through the fluctuation-dissipation theorem as $  \eta_{\parallel,\perp} = 2 M_{Q}T \kappa_{\parallel,\perp} $. 
Therefore, the drag force exerting on the heavy quarks is stronger
in the transverse direction, which generates a momentum anisotropy $ v_2 $ 
of the open heavy-flavor mesons in the final state (see Ref.~\cite{Fukushima:2015wck} for more discussions). 

\begin{table}[t]
  \caption{Breakdown of the contributions to the heavy-quark diffusion coefficient}
  \label{table:diffusion}
  \centering
  \begin{tabular}{lccccc}
    \hline
      && Longitudinal && Transverse  \\
    \hline \hline
    Quark scatterers  && 
    $ \kappa_\para^{\rm quark} = 0 $  && $ \kappa_\perp^{\rm quark} \sim \alpha_s^2 T \times eB \times \ln \frac{T^2}{\alpha_s eB} $ \\
   Gluon scatterers  && 
   $ \kappa_\para^{\rm gluon} \sim  \alpha_s^2 T \times T^2 \times \ln \frac{T^2}{\alpha_s eB}$  
   && $ \kappa_\perp^{\rm gluon} = \kappa_\para^{\rm gluon} 
   $\\
    \hline
  \end{tabular}
\end{table}

\section{Summary}

I discussed the relaxation dynamics in strong magnetic fields 
on the basis of the effective (1+1)-dimensional Boltzmann equation. 
The chirality conservation results in the enhancements 
of the longitudinal electrical conductivity \cite{Hattori:2016lqx,Hattori:2016cnt} 
and bulk viscosity \cite{Hattori:2017qih} by the inverse factor of the current quark mass. 
The same results can be obtained from the corresponding diagrammatic calculations~\cite{Hattori:2016cnt, Hattori:2017qih}. 
There are related progresses in computations in weaker magnetic fields 
where the higher Landau levels participate in the relaxation dynamics~\cite{Li:2017tgi, Fukushima:2017lvb}. 
I also discussed the heavy-quark diffusion dynamics~\cite{Fukushima:2015wck}. 
The key observation was that the chirality conservation prohibits 
the longitudinal-momentum exchange between a heavy quark and a massless quark scatterer in the LLL. 
Consequently, the diffusion coefficient acquires an anisotropy, 
which in turn gives rise to an origin of the momentum anisotropy of open heavy-flavor mesons.


\vspace{0.2cm}
{\bf Acknowledgements}.---
KH thanks Kenji Fukushima, Xu-Guang Huang, Shiyong Li, Dirk H. Rischke, 
Daisuke Satow, Ho-Ung Yee, and Yi Yin for fruitful collaborations. 
The research of KH is supported by China Postdoctoral Science Foundation
under Grant Nos.~2016M590312 and 2017T100266. 

%

\bibliographystyle{elsarticle-num}


 \biboptions{sort&compress}

\end{document}